\documentclass[aps,prl,twocolumn,showpacs,preprintnumbers,amsmath,superscriptaddress,amssymb,floats,nofootinbib]{revtex4}

\usepackage{graphicx}
\usepackage{dcolumn}
\begin{document}

\def\Journal#1#2#3#4{{#1} {\bf #2}, #3 (#4)}
\def\NCA{\rm Nuovo Cimento}
\def\NIM{\rm Nucl. Instrum. Methods}
\def\NIMA{{\rm Nucl. Instrum. Methods} A}
\def\NPB{{\rm Nucl. Phys.} B}
\def\PLB{{\rm Phys. Lett.}  B}
\def\PRL{\rm Phys. Rev. Lett.}
\def\PRD{{\rm Phys. Rev.} D}
\def\PRC{{\rm Phys. Rev.} C}
\def\ZPC{{\rm Z. Phys.} C}
\def\JPG{{\rm J. Phys.} G}
\def\st{\scriptstyle}
\def\sst{\scriptscriptstyle}
\def\mco{\multicolumn}
\def\epp{\epsilon^{\prime}}
\def\vep{\varepsilon}
\def\ra{\rightarrow}
\def\ppg{\pi^+\pi^-\gamma}
\def\vp{{\bf p}}
\def\ko{K^0}
\def\kb{\bar{K^0}}
\def\al{\alpha}
\def\ab{\bar{\alpha}}
\def\be{\begin{equation}}
\def\ee{\end{equation}}
\def\bea{\begin{eqnarray}}
\def\eea{\end{eqnarray}}
\def\CPbar{\hbox{{\rm CP}\hskip-1.80em{/}}}
\title{\large \bf The Generalized Counting Rule and Oscillatory Scaling}
\author{D.~Dutta, H.~Gao \\
{\it Triangle Universities Nuclear Laboratory and \\
Department of Physics, Duke University, Durham, NC 27708, USA}}
\begin{abstract}
We have studied the energy dependence of the $pp$ elastic scattering 
data and the pion-photoproduction data at 90$^\circ$ c.m. angle in 
light of the new generalized counting rule derived for exclusive processes. 
We show that by including the helicity-nonconserving amplitudes and their 
interference with the Landshoff amplitude, we are able to reproduce the 
energy dependence of all the $pp$ elastic cross-section and spin-correlation (A$_{NN}$) data available above the resonance region. The pion-photoproduction 
data can also be described by this approach, however, data with much finer energy spacing is needed to confirm the oscillations about the scaling 
behavior. This study strongly suggests an important role for helicity-nonconserving amplitudes related to quark orbital angular momentum and for the interference of these amplitudes with the Landshoff amplitude at GeV energies.   
\end{abstract}


\maketitle
The transition between perturbative and non-perturbative regimes 
of Quantum Chromo Dynamics (QCD) is of long-standing interest in nuclear and particle physics. Exclusive 
processes play a central role in studies trying to map out this transition.
The differential cross sections for many exclusive reactions \cite{white}
at high energies and large momentum transfers appear to obey dimensional 
scaling laws~\cite{brodsky} (also called quark counting rules). In recent 
years, the onset of this scaling behavior has been observed at a hadron transverse momentum of $\sim$ 1.2 (GeV/c) in deuteron 
photo-disintegration \cite{schulte,rossi} and in pion photoproduction from nucleon~\cite{zhu}.
On the other hand, these models also predict hadron helicity conservation in 
exclusive processes~\cite{hhc}, and experimental data in similar energy and momentum 
regions tend not to agree with these helicity conservation selection 
rules~\cite{krishni}. Although contributions from non-zero parton 
orbital angular momenta are power suppressed, as shown by Lepage and 
Brodsky~\cite{lepage}, they could break hadron helicity conservation 
rule~\cite{gousset}. Interestingly recent re-analysis of quark orbital angular 
momenta seems to contradict the notion of power suppression~\cite{rj_hh}. 
Furthermore, Ref~\cite{isgur} argues that 
non-perturbative processes could still be important in some kinematic 
regions even at high energies. Thus the transition between the 
perturbative and non-perturbative regimes remains obscure and makes it 
essential to understand the exact mechanism governing the early onset of 
scaling behavior.

Towards this goal, it is important to look closely at claims of agreement 
between the differential cross section data and the quark counting rule prediction.
Deviations from the quark counting rules have been found in exclusive 
reactions such as elastic proton-proton ($pp$) scattering~\cite{ppdata,hendry}. In fact, the re-scaled 90$^\circ$ center-of-mass $pp$ elastic scattering 
data, $s^{10}{\frac{d\sigma}{dt}}$ show substantial oscillations about the 
power law behavior. Oscillations are not restricted to the $pp$ elastic scattering channel; they are seen in elastic $\pi p$ fixed angle 
scattering~\cite{pidata} and hints of oscillation about the $s^{-7}$ scaling have also been reported in the recent data~\cite{zhu} from Jefferson Lab (JLab) on photo-pion production above the resonance region. In addition to 
violations of the scaling laws, spin correlations in polarized $pp$ elastic 
scattering also show significant deviations from perturbative QCD (pQCD) 
expectations~\cite{crabb,correlations}. Several sets of 
arguments have been put forward to account for these deviations from 
scaling laws and the unexpected spin correlations. 
Brodsky and de Teramond \cite{brodsky_de} explain the $pp$ scattering data in terms of the opening up of  the 
charm channel and excitation of $c\bar{c}uuduud$ resonant states. 
Alternatively the deviations are said to be an outcome of the interference 
between the 
pQCD (short distance) and the long distance Landshoff 
amplitude (arising from multiple independent scattering between quark 
pairs in different hadrons)~\cite{ralston}. Gluonic radiative corrections 
to the Landshoff amplitude give rise to an energy dependent phase~\cite{sen} 
and thus the energy dependent oscillation. Carlson, Chachkhunashvili, and Myhrer \cite{carlson} have also applied a similar interference concept to explain 
the $pp$ polarization data. The QCD re-scattering 
calculation of the deuteron photo-disintegration process by 
Frankfurt, Miller, Sargsian and Strikman \cite{sargsian} predicts that the
additional energy dependence of the differential cross-section, beyond the  
$\frac{d\sigma}{dt} \propto s^{-11}$ scaling, arises primarily from the 
$n-p$ scattering in the final state. In this scenario the oscillations may 
arise due to QCD final state interaction. If these predictions are correct, 
such oscillatory behavior may be a general feature of high energy exclusive 
photo-reactions. 
   
Recently, a number of new developments have generated renewed interest in this 
topic. Zhao and Close~\cite{close} 
have argued that a breakdown in the locality of quark-hadron 
duality (dubbed as ``restricted locality'' of quark-hadron duality) results in 
oscillations around the scaling curves predicted by the counting rule. They
explain that the smooth behavior of the scaling laws arise due to destructive 
interference between various intermediate resonance states in 
exclusive processes at high energies. However, at lower energies this 
cancellation due to destructive interference breaks down locally and gives 
rise to oscillations about the smooth behavior.  On the other hand, Ji {\it et al.}~\cite{ji-scaling} have derived a generalized counting rule based on a pQCD inspired model, by systematically enumerating the Fock components of a hadronic 
light-cone wave function. Their generalized counting rule for hard exclusive 
processes include parton orbital angular momentum and hadron helicity flip, 
thus they provide the scaling behavior of the helicity flipping amplitudes. The interference between the different helicity flip and non-flip amplitudes offers a new mechanism to explain the oscillations in the scaling cross-sections and spin correlations. The counting rule for hard exclusive 
processes has also been shown to arise from the correspondence between the 
anti-de Sitter space and the conformal field theory~\cite{cft} which connects superstring theory to conformal gauge theory. Brodsky {\it et al.}~\cite{brodsky_new} have used this anti-de Sitter/Conformal Field Theory correspondence or string/gauge duality  to compute the hadronic light front wave functions. This yields an equivalent generalized counting rule without the use of perturbative theory. 
Moreover, pQCD calculations of the nucleon formfactors including quark orbital 
angular momentum~\cite{Ji_ff,rj_ff} and those computed from light-front hadron 
dynamics~\cite{brodsky_new} both seem to explain the 
$\frac{1}{Q^2}$ fall-off of the proton form-factor ratio, 
$G_{E}^{p}(Q^2)/G_{M}^{p}(Q^2)$, measured recently at JLab in polarization 
transfer experiments~\cite{poltar}.

In this letter we examine the role of the helicity flipping amplitudes in the
oscillatory scaling behavior of $pp$ scattering and charged photo-pion 
production from nucleons and the oscillations in the spin correlations 
observed in polarized $pp$ scattering. We have used the generalized counting 
rule of Ji {\it et al.}~\cite{ji-scaling} to obtain the scaling behavior of 
the helicity flipping amplitudes.

It is well known that $pp$ scattering can be described by five independent 
helicity amplitudes~\cite{hel_amp_ref}. According to the dimensional as well as the generalized counting rules the three helicity-conserving amplitudes, $M(+,+ ; +,+), M(+,- ; +,-)$ and $M(-,+ ; +,-)$, have an energy dependence of $\sim 1/s^4$. On the other hand the simple constituent quark interchange 
models~\cite{hel_amp_ref} assume the two helicity flipping (nonconserving) 
amplitudes, $M(+,+ ; +,-)$($NC1$) and $M(-,- ; +,+)$ ($NC2$) to be zero. Later 
analysis by Lepage and Brodsky~\cite{lepage} have shown these amplitudes to be 
non-zero but power suppressed. The new generalized counting rule predicts their energy dependence to be $\sim 1/s^{4.5}$ and $\sim 1/s^{5}$ respectively~\cite{ji-scaling}. Thus the generalized counting, rule which includes the helicity 
flipping amplitudes and the interference between them, gives rise to 
additional energy dependence beyond the $s^{-10}$ scaling predicted by 
dimensional scaling. 

\begin{figure}[htbp]
{\includegraphics*[width=9.0cm,height=9.0cm]{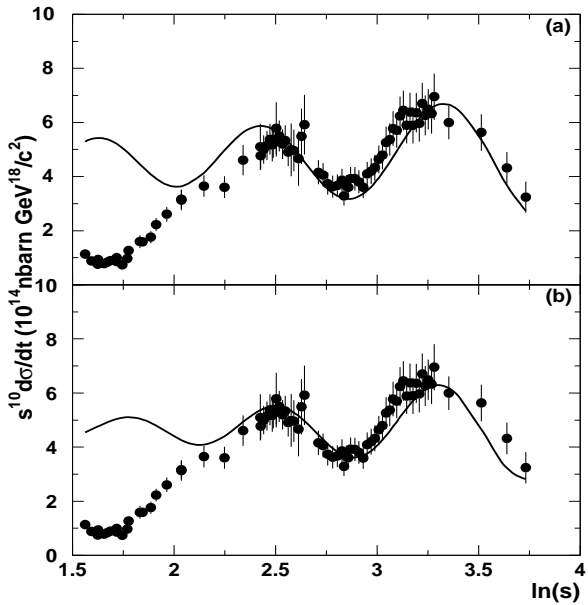}}
\caption[]{(a) The fit to $pp$ scattering data at $\theta_{cm}= 90^\circ$ of Ralston and Pire~\cite{ralston}, this fit had two parameters; the overall normalization $A_{1}$ and the arbitrary phase $\delta$. (b) The same data fitted with the new more general parametrization of the Landshoff amplitude, this fit includes the 3 additional parameters $b_1, c_1$ and $d_1$ mentioned in Eq.~\ref{kundu_eq}. The data are from Ref.~\cite{ppdata}} 
\label{ralston_fit}
\end{figure}

In addition to these short distance amplitudes, Landshoff~\cite{landshoff} 
has shown that there can be contributions from three successive 
on-shell quark-quark scattering. Although each scattering process is 
itself a short distance process, different independent scatterings can be 
far apart, limited only by the hadron size. The Landshoff amplitude also 
carries an energy dependent phase arising from gluonic radiative corrections 
which are calculable in pQCD~\cite{sen} and has a known energy dependence, 
similar to the renormalization-group evolution: $\phi(s) = \frac{\pi}{0.06}lnln(s/\Lambda_{QCD}^2)$. This effect is believed to be 
analogous to the coulomb-nuclear interference that is observed in low-energy 
charged-particle scattering. It has been shown that this energy 
dependence of the phase occurs at medium energies~\cite{botts} and 
becomes independent of energy at asymptotically high energies~\cite{botts},~\cite{mueller}. 
In Ref.~\cite{ralston}; Ralston and Pire have used the helicity-conserving amplitudes, the Landshoff amplitude with an energy dependent phase and the interference between them to reproduce the oscillations in the $pp$ scattering data at 90$^\circ$ c.m. angle (a similar method was used by Carlson {\it et. al}~\cite{carlson} to describe oscillation in the cross-section as well as the spin-correlation). They write the two amplitudes as $M = M_{S} + e^{i\phi(s) +i\delta}M_{L}$, where $M_{S}\sim 1/s^4$ represents the three helicity-conserving short distance amplitudes, $M_{L}\sim 1/s^{3.5}$ is the Landshoff amplitude and $\phi(s)$ is the energy dependent phase, $\delta$ is an arbitrary energy independent phase. By fitting to the existing $pp$ scattering data at 90$^\circ$ c.m. angle, they find that the ratio of $M_{L}$ to $M_{S}$ is 1:0.04 for an energy dependent phase given by $\phi(s) = \frac{\pi}{0.06}lnln(s/\Lambda_{QCD}^2)$, where $\Lambda_{QCD}$ = 100 MeV. It has been argued 
that the asymptotic leading limit used to calculate this energy dependence
phase of the Landshoff amplitude is not entirely valid~\cite{kundu} and thus the Landshoff term is better parametrized as,
\begin{eqnarray}
\label{kundu_eq}
 M_{L} & = & b_j s^{-3.5}\frac{e^{ic_j[lnln(s/\Lambda_{QCD})]+i\delta_j}}{[\log(s)]^{d_j}},
\end{eqnarray} 
where $b_j$, $c_j$, $d_j$ and the energy independent phase $\delta_j$ are now 
parameters which are not exactly calculable. Fig.~\ref{ralston_fit}a shows 
the fit of Ref.~\cite{ralston} compared to the world data, and 
Fig.~\ref{ralston_fit}b is a fit using the more general parametrization of 
the Landshoff described above. Both these fits deviate drastically from the data at $s<$ 10 GeV$^2$ and are not sensitive to the different parameterizations of the Landshoff amplitude. Since the Landshoff amplitude is expected to be significant only at high energies, it is not unreasonable that the above formalism  does not describe the data at low energies.    

As the interference between the Landshoff and the short distance amplitudes fail to describe the data at low energies, it is possible that the helicity flip amplitudes and their interference may play an important role at these energies. 
The helicity flip amplitudes arising from the parton orbital angular momentum 
are non-negligible when the parton transverse momentum can not be neglected 
compared with the typical momentum scale in the exclusive processes at 
relatively low energies. Thus one would expect the helicity flip amplitudes to be a significant contribution to the cross-section at low energies. Moreover, 
the generalized counting rule of Ji~{\it et al.}~\cite{ji-scaling} predicts a much faster 
fall-off with energy for the helicity flip amplitudes as expected. 
We have refitted the world data by including the two helicity-nonconserving 
amplitudes according to the generalized counting rule of Ji~{\it et al.}~\cite{ji-scaling}. The 
two different forms for the energy dependence of the phase in the Landshoff amplitude, described above, were employed in the fits to examine their sensitivity
to them. The three helicity-conserving amplitudes combined as one amplitude and the two helicity flipping amplitudes, along with the Landshoff contributions, can be written as;
\begin{eqnarray}
\label{new_eq}
M_{HC} &=& s^{-4}(a_{1} + b_{1}s^{0.5}e^{i\phi_{1}(s)}) \nonumber\\
M_{NC1} &=& s^{-4}(a_{2}s^{-0.5} + b_{2}s^{0.5}e^{i\phi_{2}(s)}) \nonumber \\
M_{NC2} &=& s^{-4}(a_{3}s^{-1} + b_{3}s^{0.5}e^{i\phi_{3}(s)}),
\end{eqnarray}
where $\phi_{j}(s)$ is the energy dependent phase. Two different forms for the phase $\phi_{j}(s)$ were used in our fits; $\phi_{j}(s)=\frac{\pi}{0.06}lnln(s/\Lambda_{QCD}^2) + \delta_j$ and  $\phi_{j}(s)=c_j\frac{lnln(s/\Lambda_{QCD}^2)+\delta_j}{(\log(s))^{d_j}}$. We have neglected the helicity flipping Landshoff contributions. The scaled cross-section is then given by, 
\begin{equation}
\label{new_eq2}
R = s^{10}\frac{d\sigma}{dt} \propto |M_{HC}|^2 + 4|M_{NC1}|^2 + M_{NC2}|^2,
\end{equation}
The factor of four associated with the $NC1$ helicity flipping amplitude 
arises because of the two possible configurations of this single spin flip amplitude~\cite{hel_amp_ref}.

\begin{figure}[htbp]
{\includegraphics*[width=9.0cm,height=9.0cm]{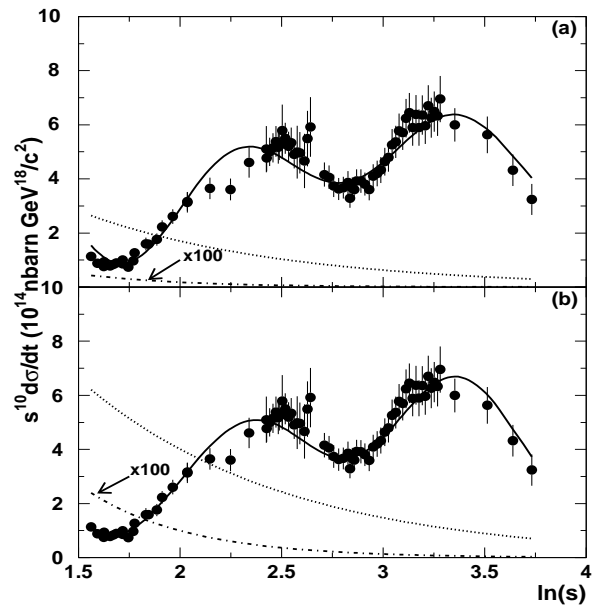}}
\caption[]{(a) The fit to $pp$ scattering data at $\theta_{cm}= 90^\circ$ when helicity flip amplitudes are included as described in Eq.~\ref{new_eq}. 
The parameters for the energy dependent phase was kept same as the earlier fit of Ralston and Pire~\cite{ralston}. The solid line is the fit result, the dotted line is contribution from the helicity flip term $\sim s^{-11}$, the dot-dashed line is contribution from the helicity flip term $\sim s^{-12}$. The $\sim s^{-12}$ contribution has been multiplied by 100 for display purposes.(b)The same data fitted to the form described in Eq.~\ref{new_eq} but with the new more general parametrization of the Landshoff amplitude which includes the 3 additional parameters per term, $b_j, c_j$ and $d_j$ ($j$=1,2,3) as mentioned in Eq.~\ref{kundu_eq}.} 
\label{newfit}

\end{figure}

 Fig~\ref{newfit} shows the results of our fit and also shows the explicit 
contributions from the $s^{-11}$ and $s^{-12}$ term for this approach. The value of $\Lambda_{QCD}$ was fixed at 100~MeV for all fits. This new fit is 
in much better agreement 
with the data. The helicity flip amplitudes (mostly the term $\sim s^{-4.5}$) are significant at low energies and seem to help in describing the data at low energies. It is interesting to note that among the helicity flip amplitudes the one with the lower angular momentum dominates. These are very promising results 
and should be examined for other reactions.

\begin{figure}[htbp]
{\includegraphics*[width=9.0cm,height=7.0cm]{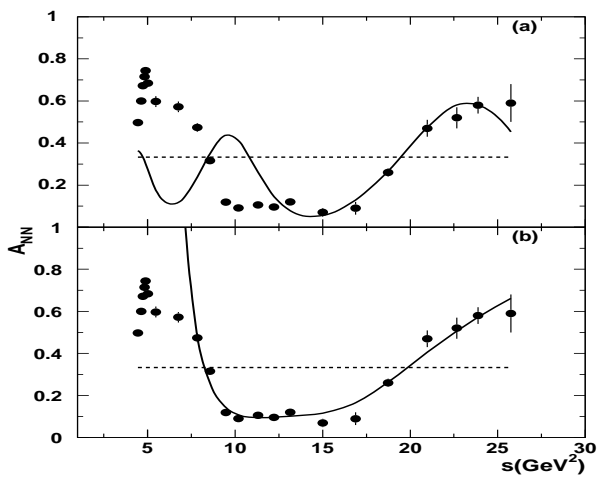}}
\caption[]{(a) The fit to $A_{NN}$ from polarized $pp$ scattering data 
at $\theta_{cm}= 90^\circ$ with the  helicity non-flip and Landshoff 
amplitudes only. (b) Fit to the same data when the helicity flip amplitudes are included. The data are from Ref.~\cite{crabb, correlations}. The solid line is the fit and the dashed line is the expectation assuming hadron helicity conservation.} 
\label{ann_fig}
\end{figure}

As mentioned earlier the $A_{NN}$ spin-correlation in polarized $pp$ elastic 
scattering also shows large deviations~\cite{correlations} from the expectations of pQCD (assuming hadron helicity is conserved). In terms of the helicity amplitudes $A_{NN}$ is given by~\cite{hel_amp_ref};
\begin{eqnarray}
\label{ann_eqn}
R A_{NN} &=& 2{\mbox{Re}}[M^*(++;++)M(--;++)] \nonumber \\ 
         &+& 2{\mbox{Re}}[M^*(+-;+-)M(-+;+-)] \nonumber \\
         &+& 4|M(++;+-)|^2, 
\end{eqnarray}
where $R$ has been defined in Eq.~\ref{new_eq2}.  
At  $\theta_{cm}= 90^\circ$ the ratio of the three helicity non-flip 
amplitudes is $2:1:1$~\cite{hel_amp_ref}. Taking this into account we have fit the $A_{NN}$ data by including the helicity flipping 
amplitudes. Fig.~\ref{ann_fig}a shows the results for the case where the helicity flip amplitude is neglected and only the interference between short 
distance amplitude and the Landshoff amplitude is used (in this case the expression for $A_{NN}$ simplifies to $R A_{NN} = 2{\mbox{Re}}[M^*(+-;+-)M(-+;+-)]$). These results are similar to those obtained by Carlson {\it et. al}~\cite{carlson} and they described the $A_{NN}$ data at high energies but fail to 
describe the low energy data using this idea of interference between 
short distance and Landshoff terms. Fig.~\ref{ann_fig}b shows the results of our fit when the helicity flipping amplitudes are included. It is clear that 
this method is a better fit to a larger fraction of the data which includes some low energy data. This suggests that even in case of the spin correlation 
$A_{NN}$ in polarized $pp$ elastic scattering the helicity flip amplitudes play an important role at low energies ($s < $ 10 GeV$^2$).   

Recently some precision data on pion-photoproduction from nucleons above the resonance region has become available from JLab~\cite{zhu}. These data show hints of oscillation about the $s^{-7}$ scaling predicted by the quark counting rule. In pion-photoproduction from nucleons the helicity non-flip amplitudes 
has an energy dependence of $s^{-2.5}$, and there is just one helicity flip 
amplitude which according to the generalized counting rule has an energy 
dependence of  $s^{-3}$~\cite{ji-scaling}. There are no leading order Landshoff terms in 
pion-photoproduction since the initial state has a single hadron. However, the 
Landshoff process can contribute at sub-leading order~\cite{farrar2} (i.e. $\sim s^{-3}$ instead of $\sim s^{-2}$). In principle, the fluctuation of a photon into a $q\bar q$ in the initial 
state can contribute an independent scattering amplitude at sub-leading 
order. But, 
experimentally it has been shown that vector-meson dominance diffractive 
mechanism is suppressed in vector meson photoproduction at large values of 
$t$ \cite{hallb}. On the other hand such independent scattering amplitude 
can contribute in the final state if more than one hadron exist in the final 
state, as is the case in nucleon photo-pion production reactions. Thus an unambiguous confirmation of such an oscillatory behavior in exclusive 
photoreactions with hadrons in the final state at large $t$ may provide a 
signature of QCD final state interaction. 

\begin{figure}[htbp]
{\includegraphics*[width=9.0cm,height=8.0cm]{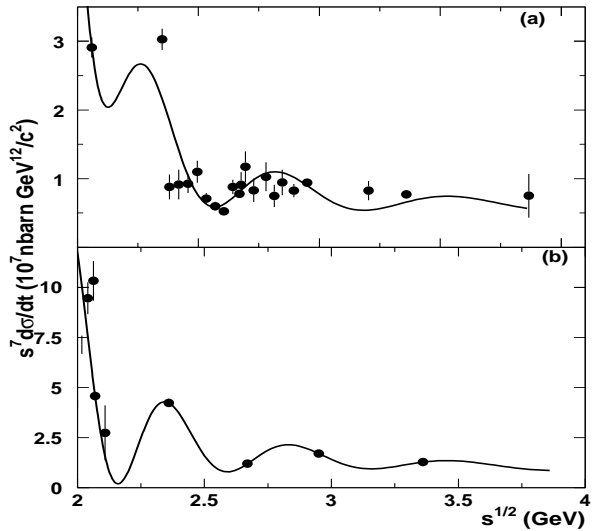}}
\caption[]{(a) The fit to $\gamma p \rightarrow \pi^+ n$ scattering data 
at $\theta_{cm}= 90^\circ$ when  helicity flip and sub-leading order Landshoff amplitudes are included (b) Fit to $\gamma n \rightarrow \pi^- p$ scattering 
data at $\theta_{cm}= 90^\circ$. The data are from Ref.~\cite{pidata, zhu}.} 
\label{pipfit}
\end{figure}

We have fit the pion-photoproduction data at $\theta_{cm}= 90^\circ$ including the helicity flip amplitude and the Landshoff amplitude at sub leading order with an energy dependent phase. The Landshoff amplitude was parametrized according to the ansatz given in Ref.~\cite{kundu}. The amplitudes for $\gamma p \rightarrow \pi^{+} n$ and $\gamma n \rightarrow \pi^{-} p$ and the respective 
Landshoff contribution to each amplitude can be written as; 
\begin{eqnarray}
\label{new_eq3}
M_{HC} &=& s^{-2.5}(a_{1} + b_{1}s^{-0.5}\frac{e^{ic_1\phi(s)+i\delta_1}}{(\log(s))^{d_1}}) \nonumber\\
M_{NC1} &=& s^{-2.5}(a_{2}s^{-0.5} + b_{2}s^{-0.5}\frac{e^{ic_2\phi(s)+i\delta_2}}{(\log(s))^{d_2}}),
\end{eqnarray}
and the scaled cross-section is given by;\\ 
$s^7\frac{d\sigma}{dt} \propto |M_{HC}|^2 + |M_{NC1}|^2$, where $\phi(s) = lnln(s/\Lambda^2)$. As seen in Fig~\ref{pipfit} the existing 
data can be fit quite well with this form. However, the data are rather 
coarsely distributed in energy and so these results are not a conclusive evidence for oscillations in pion-photoproduction. This underscores the point that a fine scan of 
energies above the resonance region is urgently needed. This is exactly the issue that will be addressed in the JLab experiment E02010~\cite{e02010} in the near future. 

We have shown that the generalized counting rule of Ji~{\it et al.}~\cite{ji-scaling} along with the Landshoff terms and associated interferences does a 
better job of describing the oscillations about the quark counting rule, in 
the $pp$ elastic scattering data at $\theta_{cm}= 90^\circ$. This is specially true in the low energy region ($s<$ 10 GeV$^2$). The contributions from helicity flipping amplitudes which are related to quark orbital angular momentum, seem to play an important role at these low energies, which is reasonable given that the quark orbital angular momentum is non-negligible compared to the
 momentum scale of the scattering process. Similarly the 
spin-correlation $A_{NN}$ in polarized $pp$ elastic scattering data can be better described by including the helicity flipping amplitude along with the 
Landshoff amplitude and their interference. The photo-pion production data 
from nucleons at large angles can also be described similarly; however, 
because of the coarse energy spacing of the data, the results are not as 
illustrative. This points to the urgent need for more data on 
pion-photoproduction above the resonance region with finer energy spacing. 
We expect that our experiment at JLab which is approved for running will help address this need in the near future.

We acknowledge fruitful discussions with X.~Ji and S.~J.~Brodsky.
This work is supported by the U.S. Department of Energy under contract number DE-FG02-03ER41231.
  

\end{document}